\documentclass[conference]{IEEEtran}
\IEEEoverridecommandlockouts
\usepackage{cite}
\usepackage{placeins}
\usepackage{float}
\usepackage{amsmath,amssymb,amsfonts}
\usepackage{algorithm}
\usepackage{algorithmic}
\usepackage{graphicx}
\usepackage{textcomp}
\usepackage{xcolor}
\usepackage{booktabs}
\usepackage{multirow}
\usepackage{tikz}
\usepackage{titlesec}
\titlespacing*{\section}{0pt}{0ex}{0ex}
\titlespacing*{\subsection}{0pt}{0ex}{0ex}
\titlespacing*{\subsubsection}{0pt}{0ex}{0ex}

\begin{document}

\title{ \huge \emph{EdgeServing}: Deadline-Aware Multi-DNN Serving at the Edge}

\author{
\IEEEauthorblockN{
Jiahe Cao\IEEEauthorrefmark{1}, 
Xiaomeng Li\IEEEauthorrefmark{1},
Qiang Liu\IEEEauthorrefmark{1},
Tao Han\IEEEauthorrefmark{2},
Ning Zhang\IEEEauthorrefmark{3},
Weisong Shi\IEEEauthorrefmark{4}\\
\IEEEauthorrefmark{1} University of Nebraska-Lincoln,
\IEEEauthorrefmark{2} New Jersey Institute of Technology,\\
\IEEEauthorrefmark{3} University of Windsor,
\IEEEauthorrefmark{4} University of Delaware,
\\
\IEEEauthorrefmark{1} \{jcao10, xli88, qiang.liu\}@nebraska.edu,
\IEEEauthorrefmark{2} tao.han@njit.edu,
\IEEEauthorrefmark{3} ning.zhang@uwindsor.ca,
\IEEEauthorrefmark{4} weisong@udel.edu
\vspace{-0.25in}
}

}

\maketitle

\begin{abstract}
As edge computing expands, serving multiple deep neural network (DNN) models on a single shared GPU has become a common yet challenging scenario, where each scheduling decision affects the tail latency of all concurrent queues. 
Existing schedulers rely on local heuristics and fail to capture this global impact, while GPU spatial-sharing approaches sacrifice latency predictability.
In this paper, we propose EdgeServing, a deadline-aware multi-DNN serving system for edge devices. 
EdgeServing adopts time-division GPU sharing with early-exit inference for high inference predictability, and introduces a stability score to quantify how each candidate scheduling decision impacts the future queue status. At runtime, it cohesively selects the model, exit point, and batch size to minimize predicted system-wide SLO impact.
Experimental results on multiple hardware platforms show that EdgeServing consistently outperforms representative baselines in both SLO violation ratio and P95 latency, enabled by early-exit mechanism, which expands the scheduling action space under tight latency constraints.

\end{abstract}

\begin{IEEEkeywords}
Multi-DNN Serving, Deadline-Aware Scheduling, Edge Computing
\end{IEEEkeywords}

\section{Introduction}
\label{sec:introduction}


Deep neural networks (DNNs) are increasingly deployed on edge devices for latency-sensitive tasks such as autonomous driving, augmented reality, and real-time video analytics~\cite{edge_ai_survey}. Unlike cloud data centers with abundant GPU resources, edge servers typically operate with a single GPU that must serve multiple DNN models concurrently~\cite{edge_computing_survey}, e.g., object detection, lane segmentation, and traffic sign classification.

In multi-DNN serving systems, GPU resources can be shared either spatial-division or time-division. 
Spatial-division sharing (e.g., GPU multi-process service or kernel-level co-location~\cite{orion, paella}) allows concurrent execution with high resource utilization.
But it introduces unpredictable latency due to blackbox contention for shared caches, memory bandwidth, and compute units in GPUs, which makes it difficult to guarantee tail latency targets. 
In contrast, time-division sharing (models run in exclusive serial mode) can achieve deterministic and reproducible inference latency~\cite{clockwork}. 
Consequently, its resource utilization may be degraded when single and batch models cannot saturate the available GPU resources. 


Extensive prior works have tackled multi-DNN serving problems from a wide range of perspectives. 
Simple heuristics such as longest-queue-first and round-robin make scheduling decisions based solely on local queue state without considering their system-wide impact~\cite{clockwork}. 
Cloud-oriented serving systems (e.g., Clockwork~\cite{clockwork}, INFaaS~\cite{infaas}) rely on cross-machine routing or abundant GPU resources to isolate models, which are unavailable on single-GPU edge devices. 
Moreover, Clockwork proactively rejects requests predicted to miss their service-level objective (SLO) deadlines~\cite{clockwork}, and Symphony~\cite{symphony} drops requests under overload when its deferred batching cannot keep pace with arrivals. 
Kernel-level schedulers such as Orion~\cite{orion} and Paella~\cite{paella} enable fine-grained spatial-division GPU sharing but target datacenter-class hardware and do not provide the latency predictability required for SLO-aware scheduling at the edge. 
Overall, multi-DNN serving at the edge remains challenging given diverse hardware platforms, software runtimes, system architectures, and time-varying traffic.

Early-exit mechanisms~\cite{shallow_deep} augment intermediate layers with lightweight exit heads and enable inference to terminate before executing the full model. Prior work on early exit has extensively studied the accuracy-latency trade-off, but predominantly in \emph{single}-model settings~\cite{shallow_deep, spinn, edgebert}. Existing methods make per-input exit decisions using runtime criteria such as confidence or entropy thresholds~\cite{shallow_deep, branchynet,apparate}, prediction consistency across successive classifiers~\cite{patience}, and learned gating policies~\cite{mue}. Apparate~\cite{apparate} extends early exit to single-model ML serving by releasing sufficiently confident intermediate predictions for latency reduction while allowing execution to continue to the final layer so that the full-model output can be used for continual accuracy monitoring and online adaptation.
However, these solutions target workload-level latency service level objectives (SLOs) rather than per-request deadline guarantees, and thus cannot ensure that each inference meets its deadline in multi-DNN serving systems.

In this paper, we propose \emph{EdgeServing}, a deadline-aware multi-DNN inference serving system at the edge. 
To meet stringent tail latency requirements, EdgeServing adopts time division GPU sharing and dynamically balances throughput, latency, and accuracy through adaptive early exit at runtime.
The key insight is that before committing to serve a model, the scheduler will \emph{predict} the resulting status across all queues, and then select the model whose service causes the least system-wide SLO damage. 
Specifically, EdgeServing introduces a new online scheduler that cohesively determines which service queue to serve, what batch size to use, and which early-exit points to choose at runtime, based on the offline profiling measurements.
In particular, we introduce a new stability score in the online scheduler to quantify the overall backlog severity across all queues.
By comparing stability scores of each candidate scheduling decision, scheduler evaluates the decisions' impacts on system and chooses the most stable decision.
We evaluate EdgeServing on diverse hardware platforms with multiple DNN models, and experimental results show that EdgeServing consistently outperforms existing scheduling baselines in SLO compliance, tail latency, and scalability across diverse workload configurations.

The main contributions are summarized as
\begin{itemize}
    \item We propose EdgeServing, a new deadline-aware multi-DNN serving system for edge devices by adopting the time-division GPU sharing approach.
    \item We design an online scheduler that cohesively optimizes model selection, early-exit point, and batch size at runtime, based on offline profile measurements.
    \item We design a stability score that evaluates the stability of the current service queues. It enables the scheduler to compare the predictive impact of candidate scheduling decisions.
    \item We conduct extensive experiments on diverse hardware platforms and multiple DNN models. Results show EdgeServing consistently outperforms existing baselines across diverse workloads and configurations.
\end{itemize}

\section{Background and Motivation}
\label{sec:background}


\textbf{GPU Resource Sharing.}
When multiple DNN models must be served on a single GPU, the system faces a fundamental resource sharing challenge. 
NVIDIA MPS~\cite{nvidia_mps} allows multiple processes to submit CUDA kernels concurrently, and shares streaming multiprocessors (SMs) through spatial partitioning. 
NVIDIA MIG partitions the GPU into dedicated instances with fixed resources, which eliminates contention but may underutilize hardware when request rates are low and bursty, a common scenario in edge serving. 
When neither MPS nor MIG is available, the GPU driver falls back to time-slicing, which multiplexes processes via context switches between quanta. 
The resulting overhead varies with the number of co-running processes, their memory footprints, and driver-internal scheduling, and cannot be reliably profiled in advance.


\textbf{Early-Exit Mechanism.}
Standard DNN inference incurs a fixed computational cost because all layers must be executed for every input.
Early-exit networks introduce intermediate classifier heads and allow inference to terminate at a selected exit point.
This mechanism exposes multiple operating points, where shallower exits provide lower latency but coarser predictions, and deeper exits provide higher accuracy.
Most prior studies investigate early exit from the perspective of per-model inference optimization and determine the exit point \emph{at runtime} through input-dependent criteria, rather than how these exit choices interact with queueing dynamics and deadline satisfaction when multiple DNNs share the same GPU.
Because the exit decision is coupled to each individual input and chosen at runtime, it is opaque to the system scheduler, which cannot proactively select an exit point to optimize system-wide objectives such as SLO satisfaction across multiple models.



\textbf{Challenge 1: Unpredictable Latency Under Concurrent GPU Execution.}
Concurrent GPU execution generally introduces latency variance from SM contention, cache thrashing, and context switching, which makes per-inference latency difficult to predict. Accurate latency prediction is essential for SLO-aware scheduling. Without it, the scheduler cannot reliably determine whether a given model-exit-batch combination will meet the deadline. In this paper, we therefore adopt a time-division sharing strategy, i.e., at any given time, the GPU serves a single model batch in exclusive mode. This ensures deterministic inference latency, which can be accurately characterized through offline profiling.

%
%

\textbf{Challenge 2: System-Wide Queue Coupling under Time-Division GPU Sharing.}
Under time-division sharing, when the GPU is allocated to one model, the request queues of all other models continue to accumulate. 
Existing scheduling policies (e.g., longest-queue-first) that neglect this system-wide coupling can rapidly trigger SLO violations. 
An early-exit mechanism introduces an additional decision dimension. Selecting a shallower exit point reduces per-request GPU service time, shortens the interval during which other models are blocked, and consequently lowers their accumulated queueing delay. 
Accordingly, the scheduler should jointly optimize model choice, exit depth, and batch size at each decision round, explicitly accounting for system-wide queue coupling induced by each action.

\section{System Overview}
\label{sec:architecture}

\begin{figure}[t]
\centering
\includegraphics[width=\linewidth]{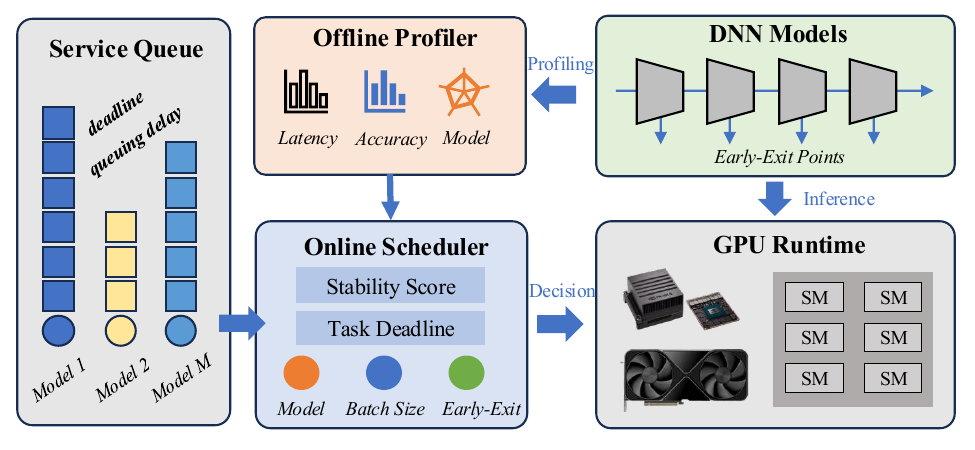}
\caption{System overview of EdgeServing.}
\label{fig:system_architecture}
\end{figure}

In Fig.~\ref{fig:system_architecture}, we show the overall architecture of EdgeServing, which includes service queues, the offline profiler, the online scheduler, and the GPU runtime.

\textbf{Service Queue.}
The system serves $M$ early-exit DNN models on a single GPU. Each model is backed by a dedicated first-in-first-out (FIFO) queue that buffers incoming inference requests. Requests arrive continuously and are enqueued regardless of the GPU's current state.

\textbf{Offline Profiler.}
Before deployment, the offline profiler benchmarks every combination of model, exit point, and batch size on the target GPU. For each combination, it records the percentile inference latency (e.g., P95) over repeated runs. The resulting profile table is loaded into the scheduler's memory at startup and remains fixed during serving. 
By running inference in time-division mode during profiling, latency measurements are deterministic and reproducible, and enable the scheduler to make accurate latency predictions at runtime. Details and results of the profiling procedure are presented in Section~\ref{sec:profiling}.

\textbf{Online Scheduler.}
The scheduler is the decision-making core of the system which runs on the CPU. 
At each scheduling step, it inspects the current state of all queues and consults the profile table to evaluate candidate scheduling actions. 
For each candidate model, it determines the appropriate exit point and batch size, then computes a stability score that predicts the system-wide impact of serving that model on all other queues' queuing time. 
The model with the lowest stability score is selected, and a batch of tasks from the corresponding queues is sent to the GPU executor. 
Comparing to simple heuristics (e.g., longest-queue-first), the scheduler explicitly accounts for cross-model contention before committing to a decision. 
The algorithmic details are presented in Section~\ref{sec:scheduler_design}.

\textbf{GPU Runtime.}
The GPU operates in time-division mode, i.e., it processes one batch at a time and is fully occupied for the duration of that inference. We will not dispatch other models to the GPU until the current batch completes. This time-division GPU sharing ensures that actual inference latency matches the profiled values, and enables the scheduler to make reliable predictions.

The \emph{EdgeServing} system operates in two phases:

\textbf{Offline Profiling Phase.}
Before the system goes online, the profiler exhaustively measures inference latency for every model-exit-batch combination on the target GPU. The profile table is generated once per hardware-model configuration and reused across all serving sessions.

\textbf{Online Serving Phase.}
Once deployed, the system enters a continuous serving loop:
\begin{enumerate}
    \item \emph{Scheduling.} The scheduler reads the current queue states (queue lengths and per-task queuing time), queries the profile table, and selects the best model-exit-batch combination.
    \item \emph{Execution.} The selected batch is dispatched to the GPU, which runs the chosen model at the chosen exit point in time-division mode. During GPU execution, no scheduling occurs, but all queues continue to accept newly arriving requests.
    \item \emph{Completion and Repeating.} Once the last batch inference is completed by the GPU executor, the scheduler immediately begins the next round based on the updated queue states. 
\end{enumerate}

\section{Model Setup and Offline Profiling}
\label{sec:profiling}

As described in Section~\ref{sec:architecture}, the offline profiler benchmarks inference latency and accuracy before deployment to provide the scheduler with accurate latency and accuracy estimates. 
This section first describes the early-exit architecture and the profiling procedure, then interprets the latency and accuracy results across all model, exit point, and batch size.

\subsection{Early-Exit Architecture}
Without loss of generality, we use the ResNet family (ResNet50, ResNet101, ResNet152) as representative backbone models in this paper.
Note that, EdgeServing is model-agnostic and readily applies to other DNNs that support intermediate exit points (e.g., VGG, DenseNet, MobileNet).

Specifically, we augment the standard ResNet architecture with early-exit classifiers to enable flexible latency-accuracy trade-offs at serving time.
The ResNet family (ResNet50, ResNet101, ResNet152) follows a common backbone structure: an initial stem (conv1, batch normalization, ReLU, max pooling) followed by four residual stages (\texttt{layer1}--\texttt{layer4}), each consisting of stacked Bottleneck blocks.
We attach a lightweight exit head after each of the first three residual stages (\texttt{layer1}, \texttt{layer2}, \texttt{layer3}), yielding three early-exit points in addition to the original network output (\texttt{final}), which follows \texttt{layer4}, global average pooling, and the fully connected classifier.
Each exit head consists of an adaptive average pooling layer followed by a single fully connected layer that maps the stage's output feature channels to the number of target classes.
When inference exits at a given point, only the backbone, included layers and target exit head are executed on the GPU, and the corresponding exit head produces the final classification output.

\subsection{Profiling Procedure}

EdgeServing conducts an offline latency profile for every combination of
model $m \in \{\text{ResNet50, ResNet101, ResNet152}\}$,
exit point $e \in \{\texttt{layer1}, \texttt{layer2}, \texttt{layer3}, \texttt{final}\}$,
and batch size $B \in \{1, \ldots, 10\}$.
Let $L(m,e,B)$ denote the profiled inference latency for configuration $(m,e,B)$ on the target GPU.
For each configuration, we run inference for hundreds of repetitions and record the average latency, form a 120-cell table (3 models $\times$ 4 exits $\times$ 10 batch sizes) that is loaded into memory at scheduler startup.
Profiles remain fixed during serving and re-profiling is only required when the GPU or model changes.
We observe that these inference latencies are stable across repeated runs (coefficient of variation $< 3\%$), which indicates the offline profiles reliably predict runtime inference latency on these platforms.

\begin{figure}[t]
\centering
\includegraphics[width=\linewidth]{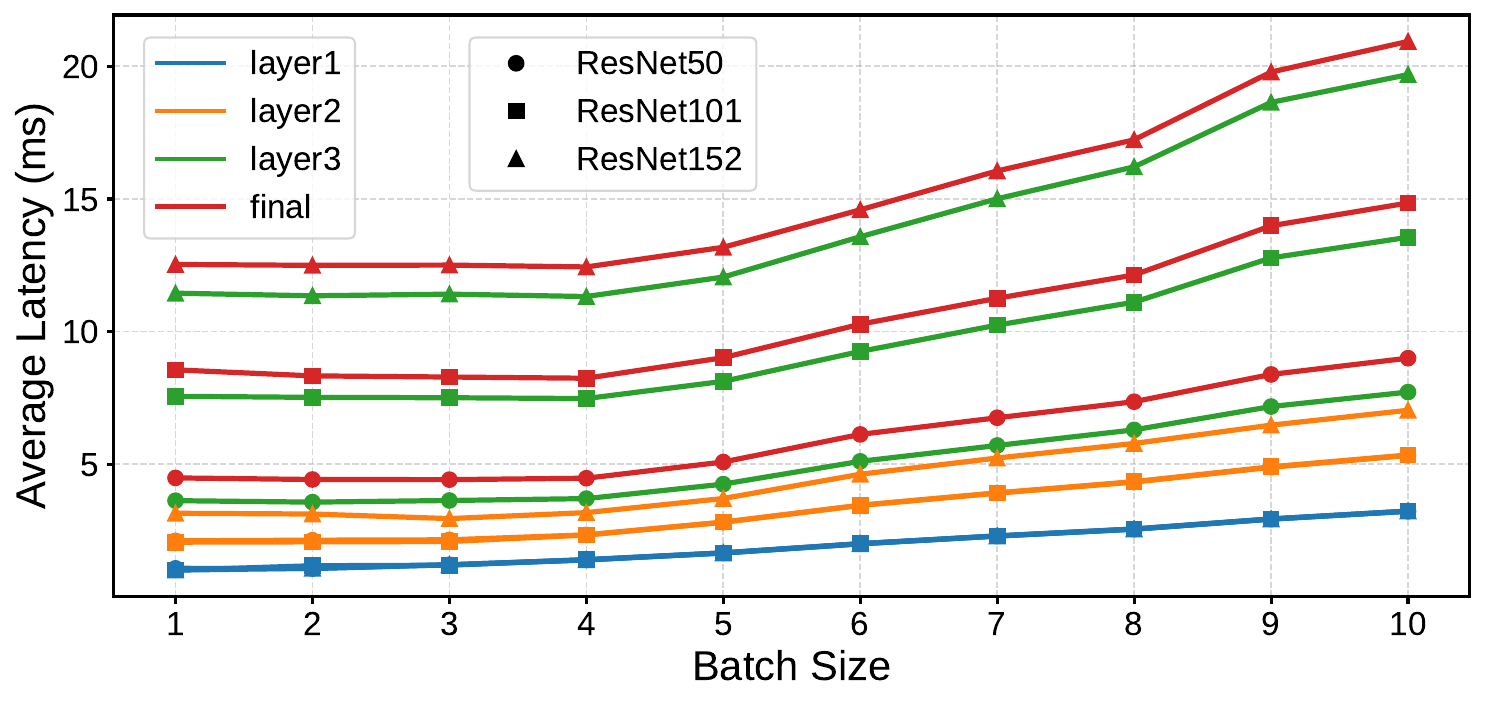}
\caption{Profiled average inference latency vs.\ batch size for all models and exit points on RTX~3080.}
\label{fig:profile_mean}
\end{figure}

\begin{table}[!t]
\renewcommand\arraystretch{1.0}
    \caption{Model accuracy (\%) at each early-exit point on CIFAR-100.} \centering
    \label{tab:accuracy} 
	\setlength{\tabcolsep}{2.5mm}{
	\begin{tabular}{lcccc}
		\toprule
		Model & layer1 & layer2 & layer3 & final \\
		\midrule
		ResNet50  & 7.6  & 12.1 & 30.8 & 74.4 \\
		ResNet101 & 7.4  & 14.5 & 54.3 & 77.9 \\
		ResNet152 & 7.3  & 17.2 & 47.4 & 78.0 \\
		\bottomrule
	\end{tabular}}
\end{table}

\subsection{Profiling Results}

To guide the online scheduler, we first characterize the profiled average latency across models, exits, and batch sizes.
Fig.~\ref{fig:profile_mean} summarizes the results and highlights three trends.
 
First, latency increases with batch size for all models and exits, but the increase is modest at small batches (when GPU cores are underutilized) and becomes more pronounced as the batch approaches GPU capacity. Overall, increasing batch size from 1 to 10 increases average latency by only about $2$--$3\times$, rather than $10\times$.
Second, deeper exits incur substantially higher latency. The \texttt{final} exit of ResNet152 is roughly 6--8$\times$ slower than its \texttt{layer1} exit at the same batch size, which confirms that early-exiting provides a meaningful latency reduction.
Third, three ResNet variants show a clear ordering (ResNet50 $<$ ResNet101 $<$ ResNet152) and the gap widens at the \texttt{final} exit, where depth differences are greatest.

These profiles feed directly into two components of the scheduler (Section~\ref{sec:scheduler_design}).
For exit-point selection, the scheduler queries the profile to find the deepest exit $e^*$ satisfying $w_{\max} + L(m, e, B) \leq \tau$ in order to ensure the chosen inference fits within the remaining SLO budget.
For stability score, the scheduler uses the same profile to predict how scheduling model $m$ will extend the queuing time of every other queue, thereby accounting for cross-model contention before committing to a decision.

\subsection{Accuracy}
To better analyze the exit points performance and scheduler trade off on latency and accuracy, we measure accuracy for each model and exit point at the same time. 
Table~\ref{tab:accuracy} summarizes the accuracy of different models and each exit points.
The reported accuracies are obtained under our training configuration without extensive hyperparameter tuning (i.e., may not be optimal). 
Note that, EdgeServing can work with any given accuracy data under different early-exit points without any further modification.



\section{Scheduler Design}
\label{sec:scheduler_design}

This section formalizes the system model and scheduling problem, then presents the online scheduler.

\begin{figure*}[t]
\centering
\includegraphics[width=1\textwidth]{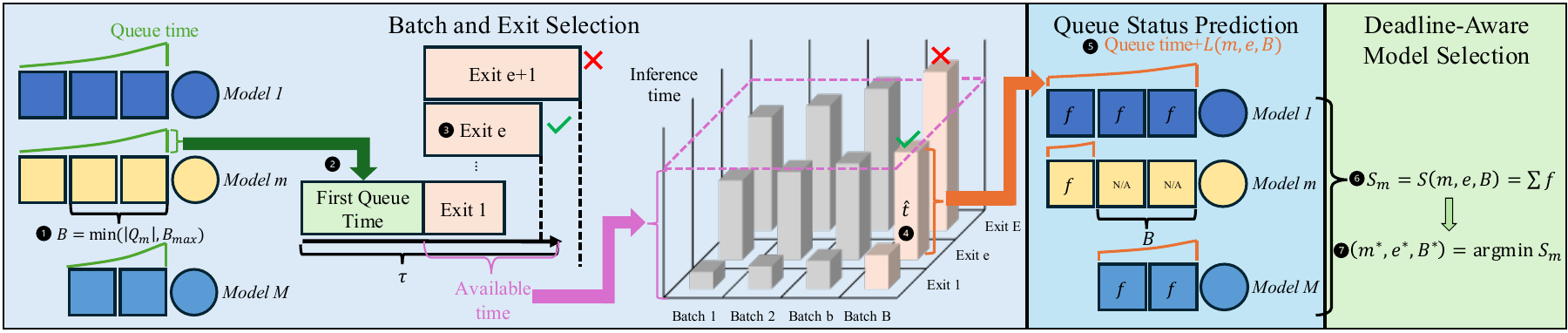}
\caption{Scheduler workflow of EdgeServing.}
\label{fig:scheduler}
\end{figure*}

\subsection{System Model}
We consider $M$ early-exit DNN models sharing a single GPU, each backed by a FIFO queue $Q_m$. 
Each model $m \in \mathcal{M} = \{1, \ldots, M\}$ has $E$ exit points $\{e_1, e_2, \ldots, e_E\}$ ordered from shallowest to deepest. 
The profile table $L(m, e, B)$ provides the percentile inference latency for model $m$ at exit $e$ with batch size $B$, obtained through offline profiling (Section~\ref{sec:profiling}). 
The GPU operates in exclusive single-model mode, i.e., it processes one batch at a time, and serving model $m$ at exit $e$ with batch size $B$ monopolizes the GPU, during which all other queues wait. 
At each scheduling round (i.e., when the previous batch inference completes), the scheduler jointly selects the model $m$, exit point $e$, and batch size $B$ and dispatch corresponding batch tasks to GPU for inference. 

\subsection{Problem Formulation}
Let $w_{m,i}$ denote the current queuing time of task $i$ at the service queue $Q_m$ at the current scheduling time (i.e., time elapsed since its arrival).
Assuming task $i$ at model $m$ is served at exit $e$ with batch size $B$, its total latency is
\begin{equation}
T_{m,i} = w_{m,i} + t_{m,i}\;,
\label{eq:total_latency}
\end{equation}
where $t_{m,i}$ is its actual inference latency. 

A task violates the SLO if $T_i > \tau$, where $\tau$ is the latency deadline.
Here, over a serving window, we define the aggregated SLO violation ratio as
\begin{equation}
V = {|\{i : T_i > \tau\}|}/{N_{\text{total}}},
\label{eq:slo}
\end{equation}
where $N_{\text{total}}$ is the total number of completed tasks.

The accuracy is directly determined by the exit point $e$, since earlier exits use shallower sub-networks with lower representational capacity, while the throughput is governed by the model $m$, the exit point $e$, and the batch size $B$, as different models vary in computational complexity and earlier exits reduce the per-inference cost.

Our scheduling goal is two-fold: 1) minimize $V$ over the serving horizon and 2) maximize the model throughput and accuracy.
The difficulty mainly lies in the coupling of service queues, where each scheduling decision changes all other queue states and also affects all subsequent decisions.
Since the future request arrivals are unknown, it is difficult, if not impossible, to derive an exact global optimum.

\subsection{Proposed Solution}
Here, we introduce the proposed online scheduler to address the aforementioned problem.
Fig.~\ref{fig:scheduler} shows the workflow of the online scheduler based on a \emph{one-step greedy} strategy.
First, we introduce a new metric (\emph{stability score}) to quantify the stability of all service queues according to the current queue status, e.g., queuing time and total latency requirement of tasks.
Second, we evaluate the impact of combinational model-exit-batch decisions and predict the queue status in the next scheduling round.
Third, we calculate the stability score based on predictive queue status under candidate decisions and select the decision to minimize the stability score in the next scheduling round.

\textbf{Stability Score.}
To quantify the urgency of a task $i$ (i.e., approaching the task's deadline) in queue $Q_m$, we use an exponential activation function with normalization as
\begin{equation}
f(w_{m,i}) = \min\!\left(\exp\!\left(w_{m,i}/{\tau} - 1\right),\; C\right).
\label{eq:activation}
\end{equation}
Here, we adopt the exponential form because task urgency grows nonlinearly as queuing time approaches the SLO deadline. 
A task at $w = 0.9\tau$ has much less remaining slack than one at $w = 0.5\tau$, and this difference should be reflected super-linearly in the score to incentivize the scheduler to avoid pushing near-deadline tasks further. 
We use the normalization ${w}/{\tau}-1$ because: it equals $0$ at $w=\tau$ for any SLO threshold, so $f(\tau)=\exp(0)=1$ and provides a consistent scale across different SLO settings.
The clip at $C$ prevents tasks already far beyond the SLO (e.g., $w > \tau(1 + \ln 10) \approx 3.3\tau$) from dominating the score. 
Such tasks have already violated the SLO and cannot be rescued by the current scheduling decision, so capping their weight avoids neglecting the remaining queues.



Hence, without loss of generality, we define the \emph{stability score} for all model queues as the sum of urgency of all tasks among all queues, which is 
\begin{equation}
S = \sum_{m \in \mathcal{M}} \sum_{i \in {Q}_{m}} f\!\left(w_{m,i}\right).
\label{eq:stability_score}
\end{equation}

\textbf{Batch and Exit Selection. }
To better search the model-exit-batch decision $(m, e, B)$, the scheduler first loops over all non-empty queues $Q_m, m \in \mathcal{M}$. 
Then, for each queue $Q_m$, it optimizes batch size $B$ and exit point $e$ to maximize the throughput ${B}/{L(m,e,B)}$ and accuracy, subject to the SLO constraint that no task in the selected batch violates the deadline. 
We will use the offline profiling measurement to approximate the actual inference latency, under different decisions.
First, we observe that, large batch always benefit to throughput in Fig.~\ref{fig:profile_mean}.
Hence, we determine the optimal batch size $B^*$ as
\begin{equation}
    B^*=\min(|Q_m|,B_{\max}),
\end{equation}
where $|Q_m|$ is the current length of the queue $Q_m$ and $B_{\max}$ is the maximum allowed batch size.
Second, we denote $w_{m,\max}$ as the maximum queue time among all queue times in $Q_m$. 
Given the optimal batch size $B^*$, we further formulate the subproblem of determining the early-exit point as 
\begin{align}
e^*=\arg\max_{e\in\mathcal{E}} \; e
\quad
\text{s.t.}\quad
w_{m,\max}+L(m,e,B^*) \le \tau.
\label{eq:batch_exit_select2}
\end{align}

Here, satisfying the constraint $w_{m,\max}+ L(m,e,B^*) \le \tau$ will guarantee that no task will violate its SLO after the GPU completes the chosen inference.
As a result, we can efficiently solve the above problem by linearly searching all the candidate early-exit points.
Overall, we can derive the optimal $(e^*,B^*)$ for each queue $Q_m$.

\textbf{Queue Status Prediction.}
Next, we predict the queue states for the next scheduling round under the assumption that the model queue $Q_m$ is selected with optimal early-exit point and batch size $(e^*, B^*)$. Because future task arrivals are unknown, they are excluded from the prediction. Based on the predicted inference time $L(m, e^*, B^*)$, the queuing time $w$ of all tasks is updated as follows.
\begin{itemize}
    \item In $Q_m$, the $B^*$ tasks being served are removed from consideration (they will complete).
    \item For each remaining task in $Q_m$ (positions $B+1, \ldots, |Q_m|$), if any, their queuing time is increased by $L(m, e^*, B^*)$.
    \item For all tasks in other queues $Q_{m'}$ ($m' \neq m$), their queuing time is increased by $L(m, e^*, B^*)$, since they cannot be served until the current batch is completed.
\end{itemize}


\textbf{Deadline-Aware Model Selection.}
With the above predictive queue status, we will calculate the stability score for all possible model queues $Q_m, m \in \mathcal{M}$. 
Finally, the scheduler selects the optimal model queue by 
\begin{equation}
    m^* = \arg \min_{m \in \mathcal{M}} S_m,
\end{equation} 
where $S_m$ is calculated according to Eq.~\ref{eq:stability_score}.
It should be noted the $B$ chosen tasks in $Q_m$ are excluded when calculating $S_m$ because they are removed from queues.
Since we derived the optimal early-exit point and batch size $(e^*, B^*)$ for each model queue, we then obtain the final model-exit-batch decision $(m^*, e^*, B^*)$.

\begin{algorithm}[t]
\renewcommand{\algorithmicrequire}{\textbf{Input:}}
\renewcommand{\algorithmicensure}{\textbf{Output:}}
\caption{Online Scheduling}
\label{alg:scheduler}
\begin{algorithmic}[1]
\REQUIRE Model set $\mathcal{M}$; FIFO queues $\{Q_m\}_{m\in\mathcal{M}}$; latency profile table $L(m,e,B)$; SLO deadline $\tau$; $B_{\max}$
\ENSURE Scheduling decision $(m^*, e^*, B^*)$
\STATE Calculate the current queuing time of all tasks $w_{m,i}, \forall m, i$
\FOR{each $m \in \mathcal{M}$ with $|Q_m|>0$}
    \STATE $S_m \leftarrow 0$
    \STATE $B, e \leftarrow BatchExitSelect(m)$ 
    \STATE $QueueStatusPredict(Q)$
    \STATE $S_m \leftarrow Deadline\textbf{-}AwareModelSelect(Q)$
\ENDFOR

\STATE $m^* \leftarrow \arg\min\limits_{m:|Q_m|>0} S_m$
\STATE $e^* \leftarrow e_{m^*}$,\;\; $B^* \leftarrow B_{m^*}$
\RETURN $(m^*, e^*, B^*)$
\end{algorithmic}
\end{algorithm}


\subsection{The Algorithm}
We summarize the online scheduling 
in Algorithm~\ref{alg:scheduler}.
At each round, the scheduler (i) computes the current queuing time of each queued task, 
(ii) evaluates each non-empty model queue as a candidate by setting $B=\min(|Q_m|,B_{\max})$ and selecting the deepest feasible exit $e$ via profile lookup while not violating SLO, 
and 
(iii) computes the stability score $S_m$ by summing the activated predicted queuing time of all remaining pending tasks, under the hypothetical choice of the serving model queue $m$. 
Finally, it selects the model with the minimum stability score and dispatches the corresponding batch tasks with exit $e$ to GPU.

\section{Experimental Evaluation}
\label{sec:experiments}

We evaluate the proposed scheduler through a series of experiments on several GPU platforms, and compare it with multiple baseline scheduling policies across various configurations.

\subsection{Experimental Setup}
\label{sec:setup}

\textbf{Hardware and Models.}
By default, we conduct experiments using an NVIDIA RTX~3080 as the primary evaluation hardware. For the cross-platform study in Section~\ref{sec:cross_platform}, we additionally deploy the scheduler on a GTX~1650 and an NVIDIA Jetson Orin Nano, and re-collect latency profiles independently on each device. 
To ensure reproducible latency measurements, GPU core and memory clock frequencies are locked to their maximum allowed values on all platforms before profiling and during experiments. 
We deploy three early-exit ResNet variants (ResNet50, ResNet101, and ResNet152) trained on CIFAR-100 with early-exit classifiers attached at four points: \texttt{layer1}, \texttt{layer2}, \texttt{layer3}, and \texttt{final}. 




\textbf{Traffic Model.}
We model the arrivals to each service queue as independent Poisson point processes. 
Request latency is measured in wall-clock time from task arrival to inference completion.
Each request randomly selects data from CIFAR-100 test batch.
The arrival rates follow the ratio $\lambda_{50} : \lambda_{101} : \lambda_{152} = 3 : 2 : 1$, which reflects the common scenario where lighter models receive heavier traffic.
On the RTX~3080, we sweep $\lambda_{152}$ from 20 to 240 req/s.
Traffic ranges for other platforms are adjusted to match each device's throughput capacity and are specified in Section~\ref{sec:cross_platform}.

\textbf{System Parameters.}
The default SLO threshold is $\tau = 50$~ms (RTX~3080 and GTX~1650), while the Jetson experiments use $\tau = 100$~ms to reflect its higher inference latency and lower throughput. 
The scheduler uses the P95 latency from pre-computed latency profiles for exit-point selection. 
The maximum batch size is 10 per model. Each experiment runs for 20 seconds, excludes the first 100 tasks as warmup.

\textbf{Baselines.}
We compare our scheduler against the following scheduling policies\footnote{Here, other systems like Clockwork~\cite{clockwork} and REEF~\cite{reef} are not included as direct baselines, because they are generally orthogonal to our solution.}:
\begin{itemize}
    \item \textbf{All-Final}: A non-adaptive full-inference baseline that selects the longest queue (LQF), dequeues up to $B_{\max}=10$ tasks, and always executes them at the deepest exit (\texttt{final}). 
    \item \textbf{All-Early}: A non-adaptive minimum-latency baseline that selects the longest queue (LQF), dequeues up to $B_{\max}=10$ tasks, and always executes them at the shallowest exit (\texttt{layer1}). 
    \item \textbf{Symphony}~\cite{symphony}: The current state-of-the-art SLO-aware scheduler, which delays dispatching each model's batch until the oldest queued request approaches its SLO deadline, thereby maximizing batch size for throughput but scheduling each model queue independently.
\end{itemize}

\subsection{Baseline Comparison}
\label{sec:baseline}

\begin{figure}[t]
\centering
\includegraphics[width=\linewidth]{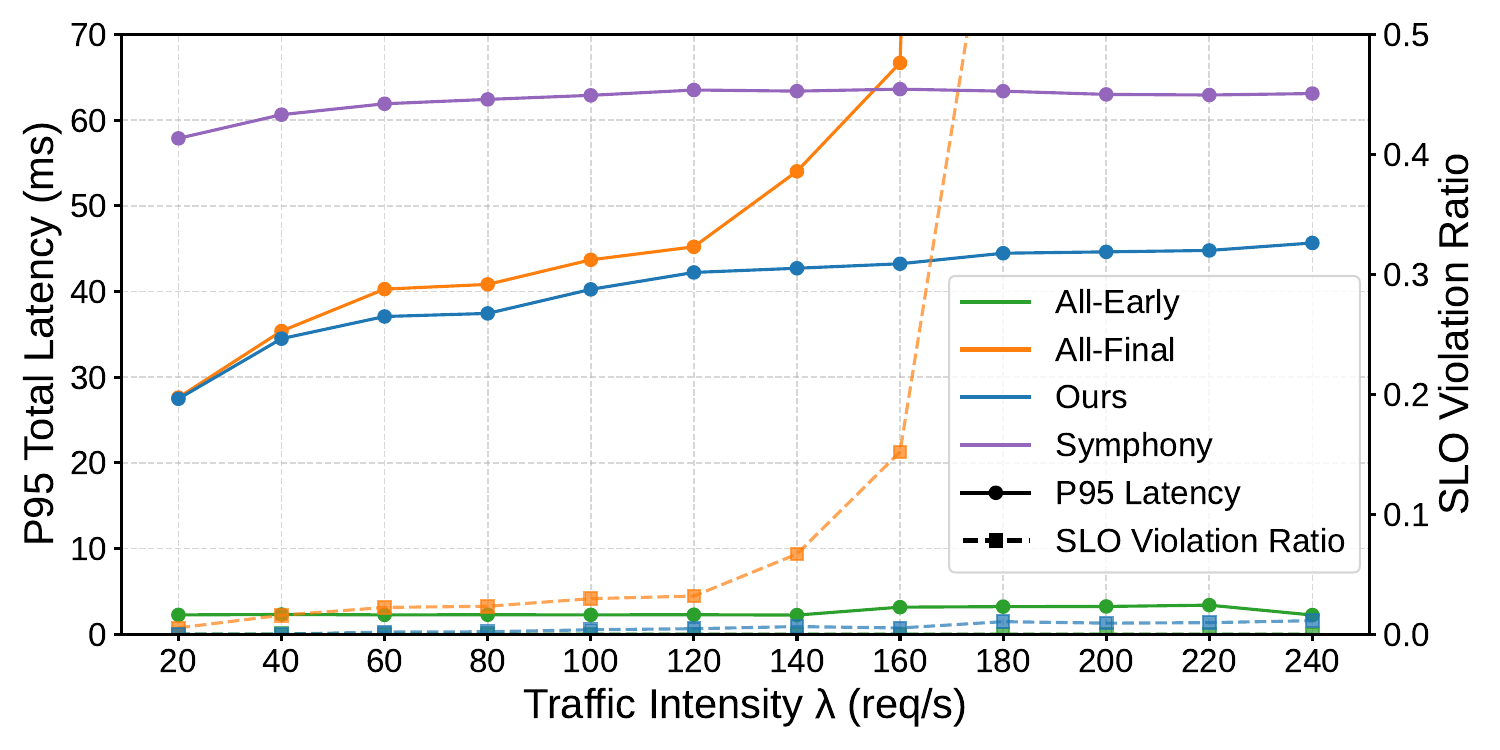}
\caption{P95 latency and SLO violation ratio vs.\ traffic intensity for schedulers.}
\label{fig:baseline_combined}
\end{figure}

\begin{figure}[t]
\centering
\includegraphics[width=\linewidth]{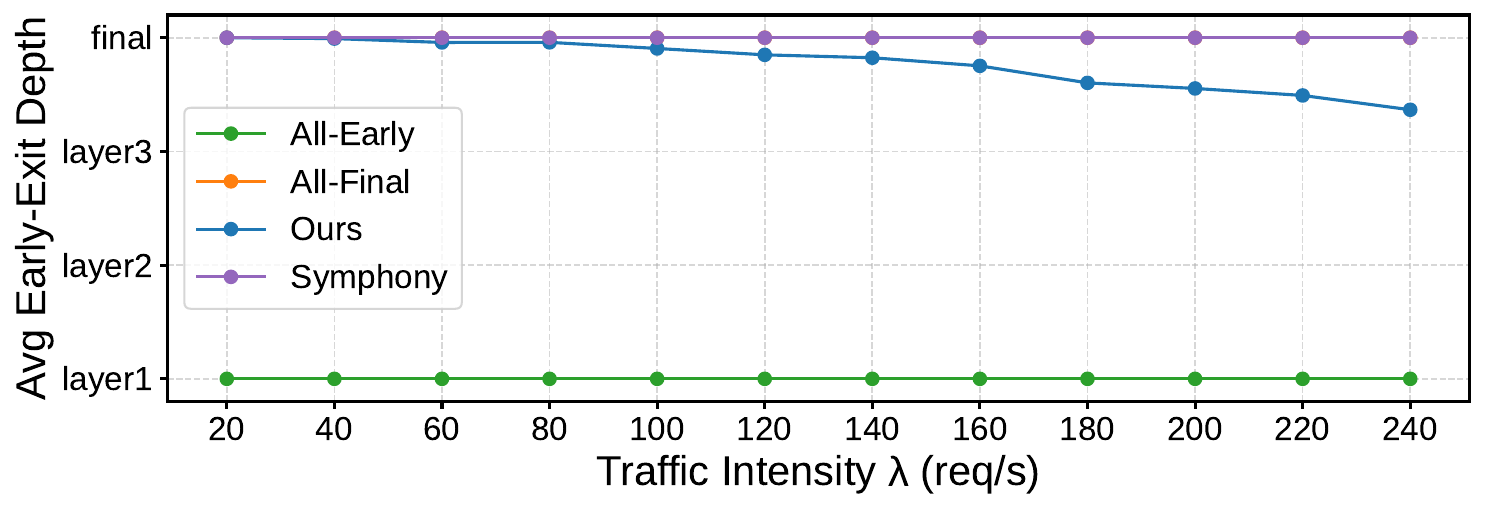}
\caption{Average early-exit depth vs.\ traffic intensity.}
\label{fig:exit_depth}
\end{figure}

Fig.~\ref{fig:baseline_combined} reports the P95 total latency (left axis) and SLO violation ratio (right axis) for all schedulers.
At low traffic, All-Final and our scheduler achieve similar P95 latency ($\sim$28\,ms), since EdgeServing can use full-depth inference when sufficient slack is available (Fig.~\ref{fig:exit_depth}).
As traffic increases, All-Final degrades sharply beyond $\lambda_{152}\approx 140$\,req/s, because full-network inference can no longer keep up with arrivals, leading to rapid queue buildup and escalating tail latency.
In contrast, our scheduler mitigates overload by switching to earlier exits when needed, and keeps P95 latency around 44--46\,ms even at $\lambda_{152}\ge 180$\,req/s.
All-Early achieves the lowest latency ($\sim$2--3\,ms) but at very low accuracy (7.4\%).
Symphony maintains a higher P95 latency of roughly 60\,ms and exhibits increasing SLO violations at higher loads, consistent with deferred batching overhead.
The violation ratio further confirms these trends that All-Final reaches 15.19\% at $\lambda_{152}=160$\,req/s, whereas our scheduler remains below 1\% across all tested intensities.

Fig.~\ref{fig:exit_depth} illustrates the adaptive exit-selection behavior. At low traffic intensity, the scheduler mostly chooses the deepest exit, whereas at higher traffic it gradually shifts toward shallower exits to meet tighter latency constraints. This trend confirms that the profile-based policy adapts exit decisions to the available latency budget.

\subsection{Accuracy and Latency Scalability}
\label{sec:accuracy_latency}
We measure system accuracy as follows. For each completed task we record the exit point used, look up the corresponding per-exit accuracy from Table~\ref{tab:accuracy}, and average over all tasks in the experiment window. 
This metric reflects the effective accuracy delivered by the system under a given traffic intensity and scheduling policy and captures the accuracy cost of using shallower exits when the SLO budget is tight. 
Since inference requests are drawn i.i.d.\ from the CIFAR-100 test set, the per-exit top-1 accuracy in Table~\ref{tab:accuracy} is an unbiased estimator of the expected accuracy for any individual request at that exit. 
Thus, the lookup-based average is a statistically reliable proxy for directly measuring per-request prediction quality.

\begin{figure}[t]
\centering
\includegraphics[width=0.9\linewidth]{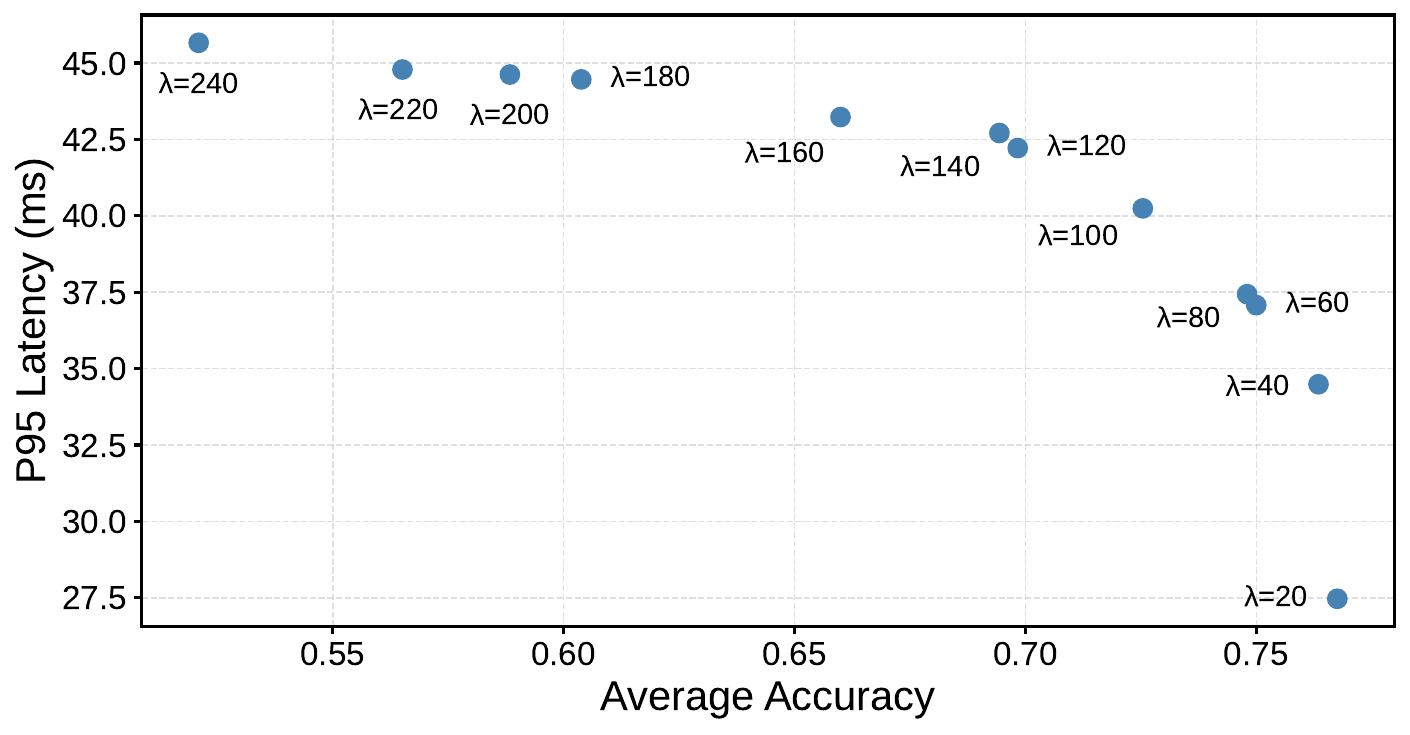}
\caption{Accuracy vs.\ P95 latency for the scheduler at different traffic intensities.}
\label{fig:acc_vs_p95}
\end{figure}

Fig.~\ref{fig:acc_vs_p95} shows the accuracy and P95 latency Pareto curve traced by the scheduler as traffic intensity varies. At $\lambda_{152} = 20$ req/s, the scheduler achieves 76.75\% average accuracy with 27.47~ms P95 latency, with the primarily \texttt{final} exit. As traffic increases to $\lambda_{152} = 100$ req/s, accuracy gracefully degrades to 72.64\% while P95 latency rises to 40.24~ms. At high traffic ($\lambda_{152} \geq 180$ req/s), the scheduler shifts to shallower exits, with accuracy settling to 60.38\% and P95 latency plateauing at 44.46~ms which remains below the 50~ms SLO. This demonstrates that the scheduler smoothly navigates the accuracy and latency trade-off space rather than experiencing abrupt degradation.

\subsection{Exit Point Configuration}
\label{sec:exit_ablation}

\begin{figure}[t]
\centering
\includegraphics[width=0.9\linewidth]{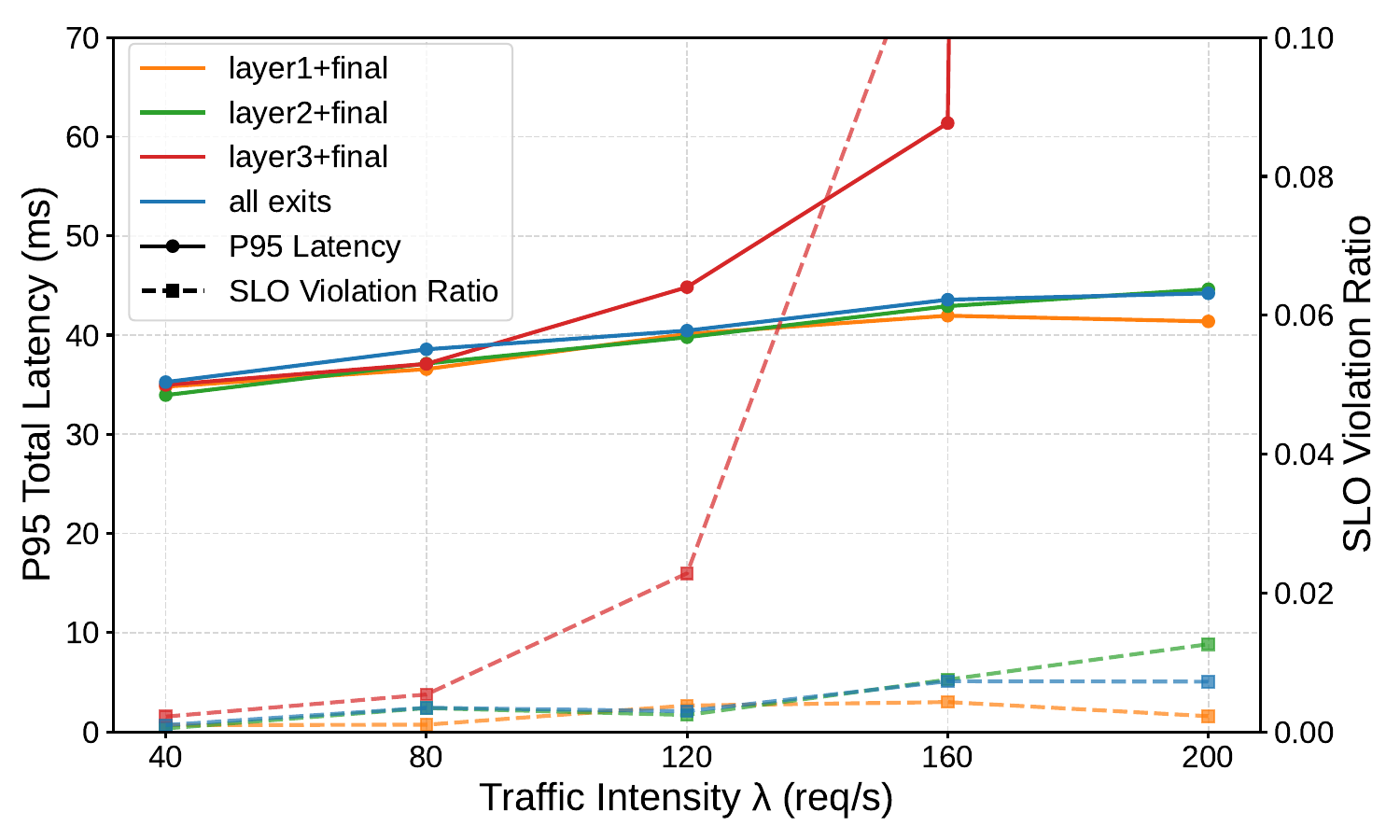}
\caption{Impact of exit point configuration on latency and SLO violation ratio.}
\label{fig:exit_ablation}
\end{figure}

We study how the choice of available exit points affects scheduler performance by testing four configurations: \texttt{layer1+final}, \texttt{layer2+final}, \texttt{layer3+final}, and \texttt{all\_exits} (\texttt{layer1+layer2+layer3+final}). Fig.~\ref{fig:exit_ablation} shows the results. \texttt{layer3+final} suffers a dramatic increase in P95 latency (over 60~ms) and violation ratio at $\lambda_{152} = 200$ req/s, because \texttt{layer3} is not fast enough to rescue tasks approaching the SLO deadline when the queue is deep. \texttt{layer2+final} shows moderate degradation at high traffic, as \texttt{layer2} provides only a partial latency reduction compared to the \texttt{final} exit. In contrast, \texttt{layer1+final} remains stable below 50~ms across all tested traffic intensities, which demonstrates that the availability of a very fast fallback exit (\texttt{layer1}) is critical for SLO compliance. Notably, \texttt{all\_exits} performs comparably to \texttt{layer1+final}, and confirms that once a sufficiently shallow exit is available, adding intermediate options provides marginal benefit.

\subsection{SLO Threshold Sensitivity}
\label{sec:slo_ablation}

\begin{figure}[t]
\centering
\includegraphics[width=0.9\linewidth]{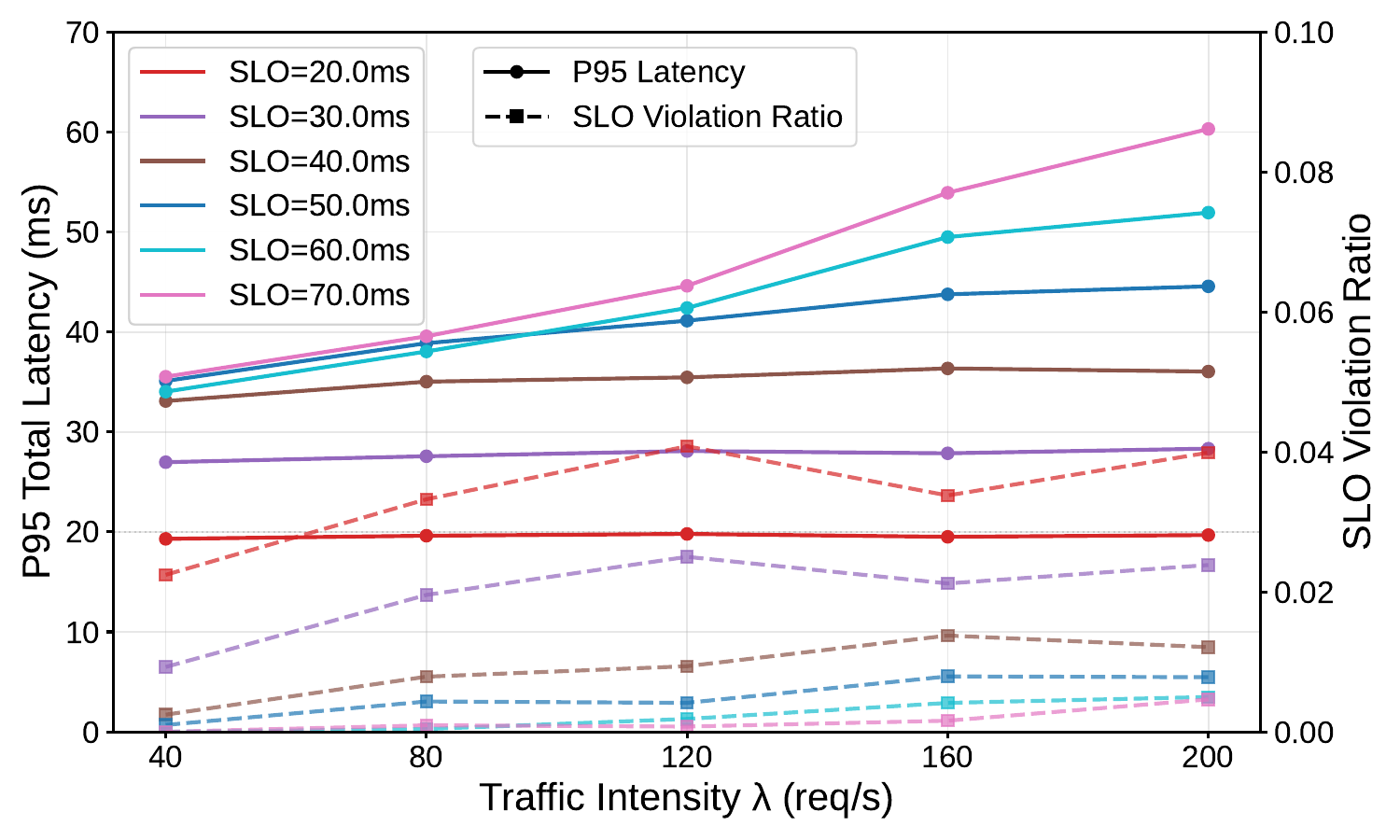}
\caption{Impact of SLO threshold on P95 latency and SLO violation ratio.}
\label{fig:slo_ablation}
\end{figure}

We evaluate the scheduler under six SLO thresholds: 20, 30, 40, 50, 60, and 70~ms. Fig.~\ref{fig:slo_ablation} presents the results.

The P95 latency scales proportionally with the SLO threshold. 
With a tight SLO of 20~ms, the scheduler aggressively uses shallow exits and keeps P95 latency at approximately 19~ms even at $\lambda_{152} = 200$ req/s. 
With a relaxed SLO of 70~ms, the scheduler uses deeper exits and P95 latency reaches 60~ms at high traffic. Notably, the P95 latency remains below the SLO threshold for each configuration at low-to-moderate traffic, which shows effective SLO compliance.

The SLO violation ratio (right axis in Fig.\ref{fig:slo_ablation}) further confirms the adaptive behavior. This result is consistent with the exit-selection trend shown earlier in Fig.\ref{fig:exit_depth}, in which tighter latency targets drive the scheduler toward shallower exits, whereas looser SLO targets permit more frequent use of deeper exits. Together, these findings indicate that the scheduler adjusts its operating point according to the target SLO.

\subsection{Model Combination}
\label{sec:model_combination}

\begin{figure}[t]
\centering
\includegraphics[width=0.9\linewidth]{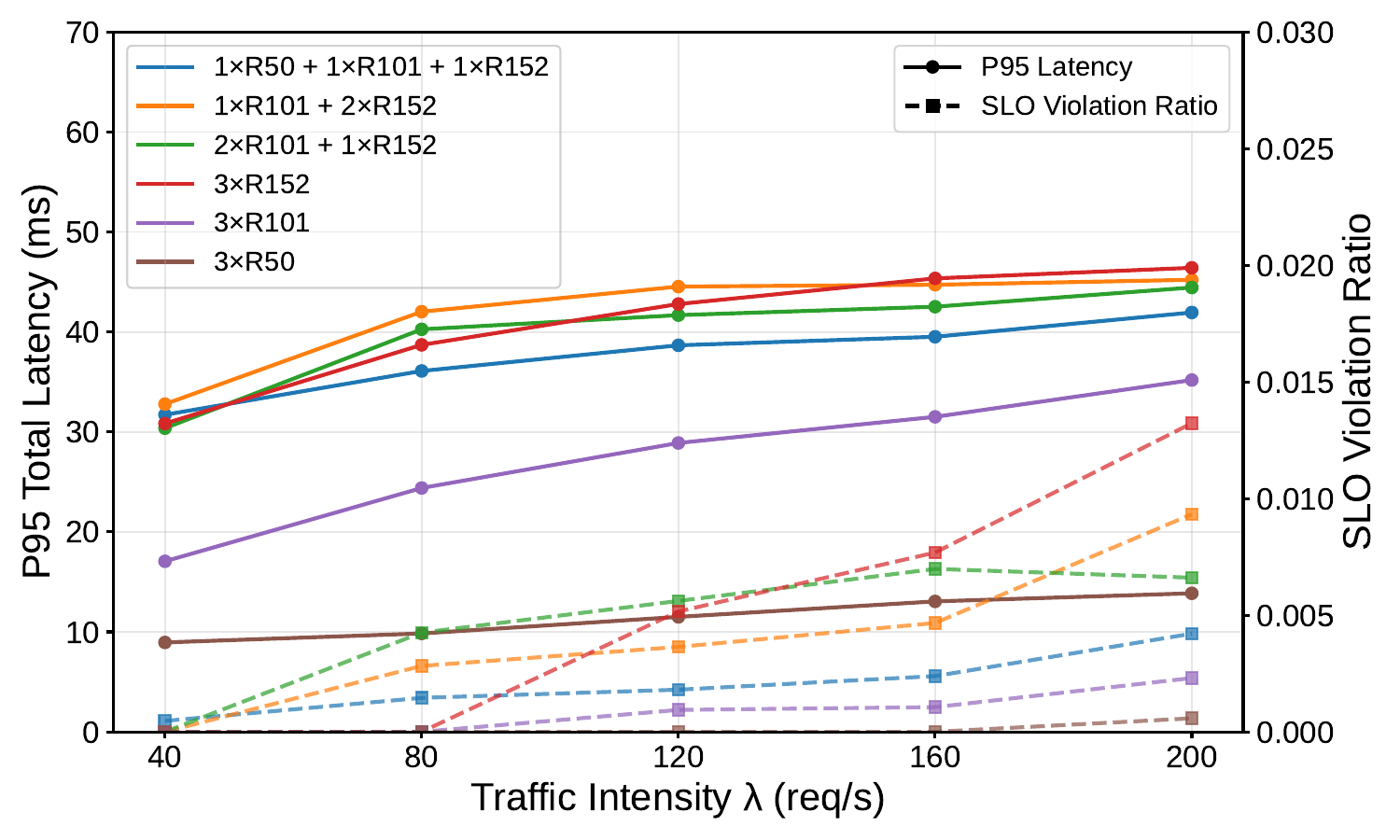}
\caption{Impact of model combination on P95 latency and SLO violation ratio.}
\label{fig:model_combo}
\end{figure}

We evaluate the scheduler under six different model deployment configurations to test its robustness to model heterogeneity. The configurations range from homogeneous deployments (3$\times$ResNet50, 3$\times$ResNet101, 3$\times$ResNet152) to heterogeneous combinations. Unlike the baseline experiments where arrival rates follow a 3:2:1 ratio, here all three model queues receive equal traffic ($\lambda_{50} : \lambda_{101} : \lambda_{152} = 1 : 1 : 1$) so that the effect of model heterogeneity is not confounded by asymmetric load. Fig.~\ref{fig:model_combo} shows the results.

The 3$\times$ResNet50 configuration achieves the lowest P95 latency and near-zero violation ratios, as ResNet50 has the smallest computational cost. The 3$\times$ResNet101 configuration shows moderate P95 latency with low violations. In contrast, configurations involving ResNet152 instances exhibit higher latency due to the heavier computational demands.

The heterogeneous baseline configuration (1$\times$ResNet50 + 1$\times$ResNet101 + 1$\times$ResNet152) achieves competitive performance, with P95 latency between 31--42~ms and violation ratios below 0.5\% across all tested traffic intensities. 
These results suggest that the scheduler can accommodate heterogeneous model costs while maintaining good latency performance and low violation rates.
This demonstrates that the scheduler effectively balances workload across models of different computational costs by predicting and minimizing cross-model contention.

\subsection{Cross-Platform Generalization}
\label{sec:cross_platform}

\begin{figure}[h]
\centering
\includegraphics[width=\linewidth]{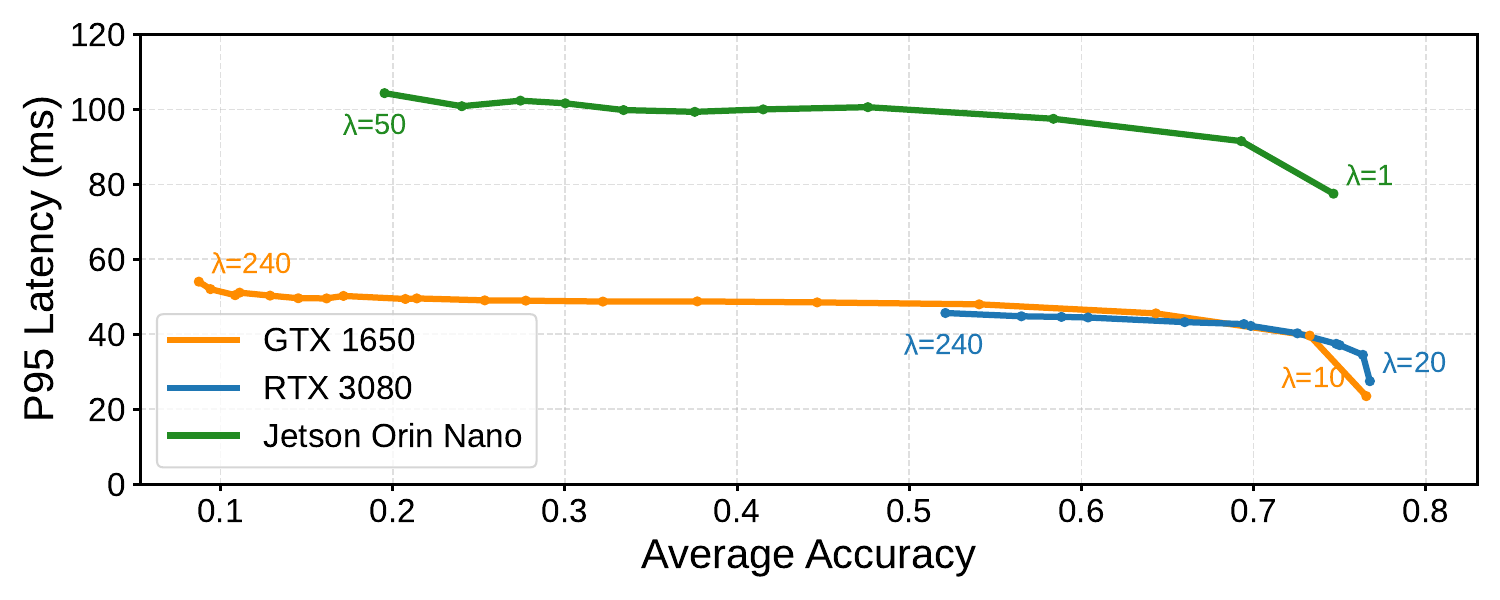}
\caption{Accuracy and latency trade-off under different hardware.}
\label{fig:cross_platform}
\end{figure}

To assess the scheduler's generalizability beyond high-end edge GPUs, we deploy the same algorithm on two additional platforms, a GTX~1650Super and an NVIDIA Jetson Orin Nano. Latency profiles are re-collected on each platform. Fig.~\ref{fig:cross_platform} shows the accuracy and P95 latency trade-off for all three platforms.


As shown in Fig.~\ref{fig:cross_platform}, on the RTX~3080 the scheduler maintains high accuracy (55--77\%) with P95 latency well within the 50~ms SLO. On the GTX~1650, higher per-inference latencies shrink the feasible deep-exit region, so the scheduler retreats to shallower exits earlier. 
At $\lambda_{152} = 240$ req/s accuracy drops to $\sim$9\% because even \texttt{layer1} cannot keep up with the arrival rate. 
On the Jetson Orin Nano ($\tau = 100$~ms), the scheduler achieves $\sim$75\% accuracy at low traffic but degrades to $\sim$20\% as the platform saturates, with P95 latency reaching 77--105~ms. 
In both weaker platforms, the accuracy drop reflects hardware throughput limits rather than scheduling deficiency and the scheduler correctly selects the best available exit but cannot compensate when GPU capacity is exceeded.

Across all three platforms, the scheduler exhibits the same qualitative behavior (deep exits at low traffic, progressive shallowing under load) with no algorithmic modification, only re-collected latency profiles.

\subsection{Ablation Study: Core Design Components}
\label{sec:ablation_eelqf}

\begin{figure}[!t]
\centering
\includegraphics[width=0.9\linewidth]{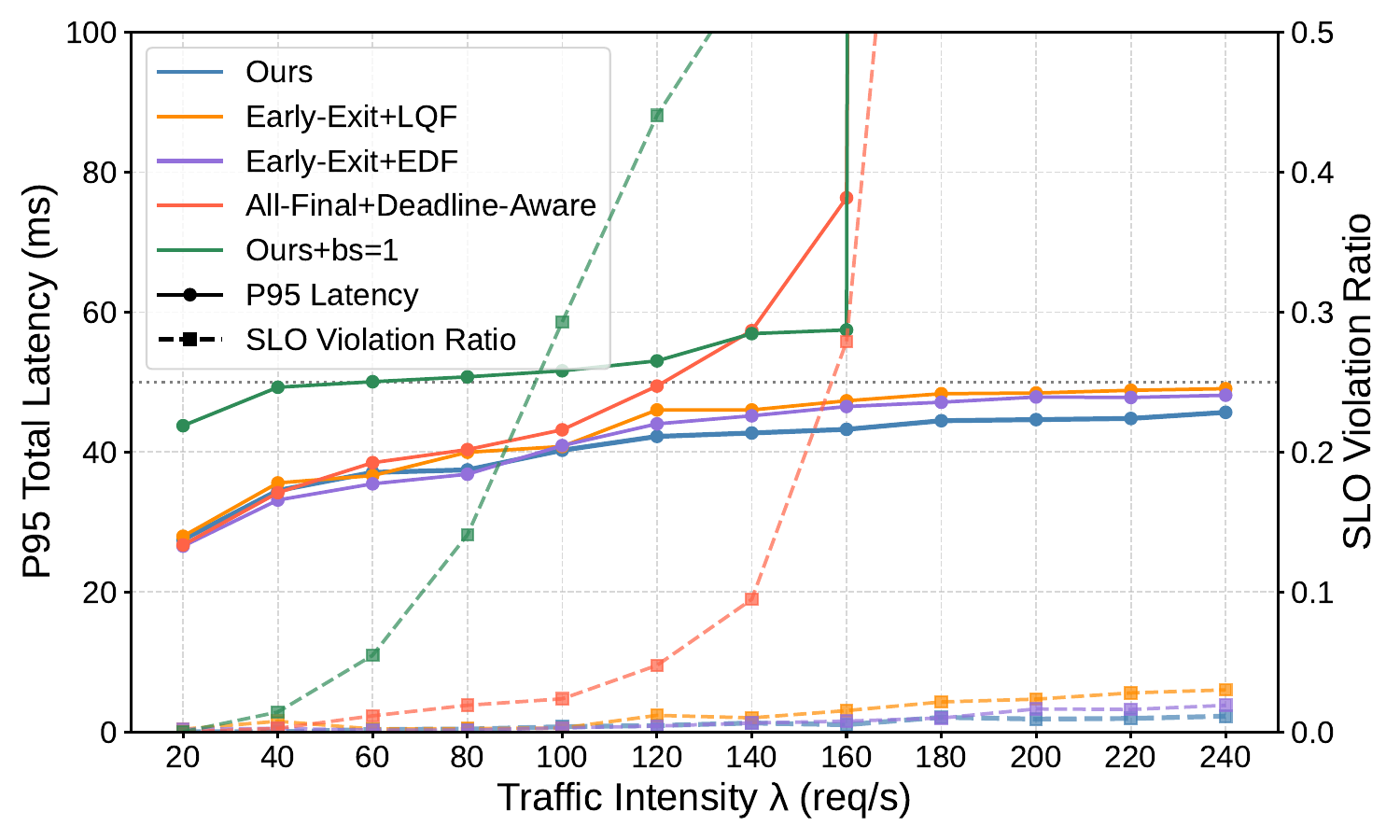}
\caption{P95 latency and SLO violation ratio for ablation study.}
\label{fig:core_ablation}
\end{figure}

To quantify the contribution of each design component, we compare the full scheduler against four ablated variants:
\begin{itemize}
    \item \textbf{Early-Exit+LQF}: retains profile-based exit selection but replaces deadline-aware model scoring with longest-queue-first (LQF), which always selects the model with the most queued tasks. This isolates the contribution of \emph{deadline-aware model selection} against a simple queue-length heuristic.
    \item \textbf{Early-Exit+EDF}: retains profile-based exit selection but replaces deadline-aware model scoring with earliest-deadline-first (EDF), which always selects the model whose oldest queued task has the least remaining SLO slack. This isolates the contribution of \emph{deadline-aware model selection} against a simpler deadline-priority rule.
    \item \textbf{All-Final+Deadline-Aware}: retains deadline-aware model scoring but disables early exit, always executing at the \texttt{final} exit. This isolates the contribution of \emph{early-exit adaptation}.
    \item \textbf{Ours+bs=1}: retains both early exit and deadline-aware scoring but fixes batch size to 1, which disables dynamic batching. It isolates the contribution of \emph{dynamic batching}.
\end{itemize}
Specifically, Fig.~\ref{fig:core_ablation} shows P95 latency and SLO violation ratio for all variants.

\textbf{Effect of Deadline-Aware Model Selection.}
At low traffic, all three model-selection strategies (LQF, EDF, stability score) converge to similar choices and perform comparably.
As traffic grows, the gap separates at $\lambda_{152} = 120$ req/s, our scheduler achieves 42.22~ms P95 latency versus 44.02~ms for Early-Exit+EDF and 46.01~ms for Early-Exit+LQF.
At $\lambda_{152} = 240$ req/s, Early-Exit+EDF reaches a 1.89\% violation ratio and Early-Exit+LQF reaches 2.99\%, while our scheduler remains below 1\%.
Early-Exit+EDF outperforms Early-Exit+LQF because deadline priority avoids the starvation that LQF can cause. However, it still falls short of our scheduler because EDF selects based on individual task urgency without predicting the system-wide impact of serving that model on other queues. The stability score accounts for this cross-model contention and allows the scheduler to avoid decisions that globally degrade queuing time even when no single queue appears urgent.

\textbf{Effect of Early-Exit Adaptation.}
All-Final+Deadline-Aware performs similarly to our scheduler at low traffic, where full-inference latency still fits within the SLO.
Similar to All-Final, the gap widens as traffic increases. The inability to reduce per-request inference time causes queue backlogs. Beyond the saturation point, P95 latency and violation ratio rise sharply at $\lambda_{152} = 120$ req/s and explode at $\lambda_{152} = 160$ req/s.
This confirms that early-exit adaptation is the primary mechanism sustaining SLO compliance under load. It also shows deadline-aware scoring alone is insufficient without a fast fallback exit path.

\textbf{Effect of Dynamic Batching.}
Ours+bs=1 incurs higher P95 latency and violation ratios than the full scheduler across all traffic intensities, as single-task execution underutilizes GPU parallelism and reduces throughput. 
Lower throughput causes queues to deepen faster and makes deadline-aware predictions less effective at preventing SLO violations under high load.

Together, all three components are necessary. Early-exit adaptation provides the latency headroom to meet the SLO, deadline-aware model selection minimizes cross-model contention, and dynamic batching maximizes GPU utilization to sustain throughput.

\section{Related Work}
\label{sec:related_work}

\subsection{DNN Inference Serving Systems}


Cloud-oriented serving frameworks such as Clipper~\cite{clipper}, TensorFlow Serving~\cite{tfserving}, and NVIDIA Triton~\cite{triton} provide dynamic batching and model management but assume abundant, largely independent GPU resources.
Clockwork~\cite{clockwork} and INFaaS~\cite{infaas} improve predictability and model selection via cross-machine routing, which is unavailable on single-GPU edge devices.
At the kernel level, Orion~\cite{orion} and Paella~\cite{paella} enable fine-grained spatial GPU sharing through CUDA interception and compiler co-design, but target datacenter-class hardware and operate at a fundamentally different granularity from our request-level scheduling.
None of these systems exploit intra-model early exits as a scheduling dimension.

\subsection{Early-Exit Networks}


Early-exit mechanisms attach auxiliary classifiers at intermediate layers, allowing inference to terminate early to trade accuracy for latency.
Existing exit criteria fall into three broad categories.
(i)~Confidence or entropy thresholds: BranchyNet~\cite{branchynet} and Shallow-Deep Networks~\cite{shallow_deep} gate each sample by its prediction confidence or entropy, as well as extensions to edge settings and modern architectures~\cite{spinn, edgebert}.
(ii)~Prediction consistency: Patience~\cite{patience} triggers an exit when consecutive internal classifiers produce unchanged predictions.
(iii)~Learned gating modules: MuE~\cite{mue} employs a lightweight decision module at each exit for input-adaptive exiting.
These works primarily focus on per-request accuracy latency trade-offs under single model scenarios, without addressing system level scheduling when multiple early-exit models share a single accelerator.

\subsection{SLO-Aware Scheduling}


Prior SLO-aware serving systems aim to meet latency targets via batching and adaptive model configuration.
Symphony~\cite{symphony} performs deadline-driven batching to improve throughput under latency constraints, while systems such as Proteus~\cite{proteus} and Cocktail~\cite{cocktail} adapt model configurations/variants to satisfy SLOs.
REEF~\cite{reef} explores GPU scheduling mechanisms (e.g., preemption) for real-time inference.
However, these approaches generally treat models in isolation or focus on model-level variant selection, and do not explicitly account for system-wide model contention when multiple model queues share a single GPU, nor do they leverage intra-model early-exit points as a scheduling degree of freedom.

\vspace{0.05in}
\section{Conclusion}
\label{sec:conclusion}
\vspace{0.05in}
We presented a new deadline-aware multi-DNN serving system on edge devices. By cohesively optimizing model selection, early-exit point, and batch size at each scheduling round, our proposed system predicts and minimizes system-wide SLO impact, which achieves low violation ratios and robust performance across diverse hardware platforms, traffic conditions, and system configurations.

\vspace{0.05in}
\section*{Acknowledgment}
\vspace{0.05in}
This work is partially supported by the US National Science Foundation under Grant No. 2321699 and No. 2426481. 

\vspace{0.1in}
\bibliographystyle{IEEEtran}
\bibliography{references}

\end{document}